
\def\bra#1{\left\langle #1\right|}
\def\ket#1{\left| #1\right\rangle}
\input lecproc.cmm
\contribution{Pion and Sigma Polarizabilities \newline
and Radiative Transitions}
\author{Murray A. Moinester @1}
\address{@1School of Physics and Astronomy \newline
Raymond and Beverly Sackler Faculty of Exact Sciences, \newline
Tel Aviv University, 69978 Ramat Aviv, Israel, \newline
e-mail: murray tauphy.tau.ac.il \newline \newline
Bulletin Board: hep-ph xxx.lanl.gov/9409463 \newline
Tel Aviv U. preprint TAUP 2204-94 \newline
Contribution to Proceedings of
Workshop on Chiral Dynamics, \newline
Massachusetts Institute of Technology, July
1994, \newline  Eds. A. Bernstein, B. Holstein.}
\abstract {Fermilab E781 plans measurements of  $\gamma$-Sigma and
$\gamma$-pion interactions using a 600 GeV beam of Sigmas and pions, and a
virtual photon target. Pion polarizabilities and radiative transitions will be
measured in this experiment. The former can test a precise prediction of
chiral symmetry; the latter for $a_1(1260) \rightarrow \pi \gamma$ is
important for understanding the polarizability. The experiment also measures
polarizabilities and radiative transitions for Sigma hyperons. The
polarizabilities can test predictions of baryon chiral perturbation theory.
The radiative transitions to the $\Sigma^*(1385)$ provide a measure of the
magnetic moment of the s-quark. Previous experimental and theoretical results
for $\gamma\pi$ and $\gamma\Sigma$ interactions are given. The E781 experiment
is described.}

\titlea{1}{Introduction}
Pion and $\Sigma$(1189) polarizabilities  and associated radiative
transitions will be measured in the Fermilab E781 SELEX experiment $[1,2]$.
Hadron electric ($\alpha$)  and magnetic ($\beta$) intrinsic polarizabilities
$[2-6]$ characterize the induced transient dipole moments of hadrons subjected
to external oscillating electric $\vec{E}$ and magnetic $\vec{H}$ fields. The
dipole moments are given by $\vec{d} = \alpha\vec{E}$ and by $\vec{\mu} =
\beta \vec{H}$.  The polarizabilities can be obtained from precise
measurements of the gamma-hadron Compton scattering differential cross
section. They probe the rigidity of the internal structure of baryons and
mesons, since they are induced by the rearrangement of the hadron constituents
via action of the  photon electromagnetic fields during scattering. For the
light charged pion, chiral symmetry leads to a precise prediction for the
polarizabilities $[7,8]$. The experimental polarizabilities therefore subject
the chiral perturbation techniques of QCD to new and serious tests. The
availability of high energy $\Sigma$ beams raises the possibility of
investigating their Compton scattering polarizabilities and radiative
transitions. The $\Sigma$ polarizabilities can test predictions of baryon
chiral perturbation theory $[9]$. The radiative transitions to the
$\Sigma^*(1385)$ can provide a valuable measure of the magnetic moment of the
s-quark $[10,11]$.

\titlea{2}{Sigma Polarizabilities and Radiative Transitions}

Electromagnetic interactions of $\Sigma$'s can be studied $[2,11]$ in FNAL
experiment E781 with high energy $\Sigma$ beams via the Primakoff interaction
of an incident $\Sigma$ with a virtual photon in the Coulomb field of a target
nucleus of atomic number Z. The $\gamma \Sigma$ Compton scattering and
associated polarizabilities are studied $[2,11]$ by detecting the final state
$\gamma$-ray  and $\Sigma$ in coincidence. The final state Sigma or other
charged particles will be measured $[1,2]$ in magnetic spectrometers, while
final state $\gamma$-rays will be measured with lead glass electromagnetic
calorimeters. The laboratory decay length  ($L = {\gamma}{\beta}c{\tau}$) for
very high energy $\Sigma$'s exceeds ten meters, so that magnetic detection is
possible. Transition radiation detectors upstream of the target and ring
Cerenkov detectors downstream will provide particle identification $[1]$. The
decay signature for $\Sigma^*$(1385) 3/2$^+$ production is clean as it decays
to $\Lambda\pi$, and the $\Lambda$ to $\pi^-$p. Such radiative transition
studies were first suggested by Lipkin $[10]$ and considered in more detail
afterwards $[2,11-15]$. The $\Sigma^0 \rightarrow \Lambda \gamma$ radiative
transition (lifetime) was measured in this way $[15]$ using a $\Lambda$ beam,
and this to date is the only precision hyperon radiative transition measured.
Because radiative transitions involve the well understood electromagnetic
process, they provide precision tests of the quark wave functions
characterizing the configurations of the low lying excited states of hadrons.

The $\Sigma^+$ (uus) and $\Sigma^-$ (dds) baryons are of interest because
electromagnetic observables are particularly sensitive to their underlying
quark structure $[11]$. The $\Sigma^+$ differs from the proton only by
replacing the $d$ quark by a strange quark which also has charge -1/3. Thus,
any difference between the electromagnetic properties of the $\Sigma^+$ and
proton can only arise from flavor SU(3) symmetry breaking. The two $\Sigma$
states are isospin mirrors of one another, symmetrically placed in the same
flavor SU(3) octet and have very similar strong interactions. However, their
electromagnetic interactions are completely different because the $\Sigma^+$,
like the nucleon and $\Xi^o$, have valence quarks of two flavors having the
opposite sign of electric charge. External electromagnetic fields therefore
exert forces in opposite directions on the two flavors and rotate spins in
opposite directions, thereby producing internal excitation. The $\Sigma^-$ on
the other hand, and also the $\Xi^-$, have three valence quarks all with
charge -1/3, and the principle effect of an external electromagnetic field is
to exert forces in the same direction on all three valence quarks and rotate
their spins in the same direction. This produces no internal excitation. We
discuss below experimental implications of this effect, first noted $[10]$ by
a selection rule involving the U-spin SU(2) subgroup of SU(3). U-spin is
conserved to all orders in any combination of  electromagnetic interactions
and strong interactions invariant under SU(3). The resulting SU(3) prediction
is that the Primakoff excitation  $\Sigma^- \rightarrow\Sigma^{*-}$ is
forbidden, while excitation of $\Sigma^+ \rightarrow \Sigma^{*+}$ is allowed.
The U-spin values of interest are U=1/2 for  $\Sigma^-$, $\Sigma^+$,
$\Sigma^+$(1385) and U=3/2 for $\Sigma^-$(1385).

  The E1 and M1 $\Sigma$ transitions are related to the intrinsic electric and
magnetic $\Sigma$ polarizabilities. We will elucidate below this connection,
which can be tested experimentally. Hadron electric ($\alpha$)  and magnetic
($\beta$) intrinsic polarizabilities characterize the induced transient dipole
moments. During $\gamma$-hadron Compton scattering, the lowest order scattering
(Thomson) is determined by the charge and magnetic moment. The next order
scattering (Rayleigh) is determined by the induced dipole moments. The Compton
cross section data determine the Compton polarizabilities $\bar{\alpha}$ and
$\bar{\beta}$, expressed here in Gaussian units of $10^{-43}$ cm$^3$. The
angular distribution formulae for low $\gamma$-ray energies in terms of
$\bar{\alpha}$ and $\bar{\beta}$ for $\gamma$p scattering are given by
Petrunkin and Lvov $[4,5]$, and similar expressions apply for the $\Sigma$
hyperons. Perturbation theory for the $\Sigma^+$, $\Sigma^-$, and proton
polarizabilities can be applied, leading to the expressions $[4]$ of
Petrunkin:
$$ \bar{\alpha} = {2\sum {{| \langle0|\vec{d_z}|n\rangle |}^2} \over
{E_n-E_0}} + {{1}\over{3}}{{e^2}\over{M}} {\langle{r}^2\rangle} = \alpha +
\Delta\alpha, \eqno (1) $$

$$ \bar{\beta} =  2\sum{{{|\langle0|\vec{\mu_z}|n\rangle|}^2}\over {E_n-E_0}} -
k{{e^2}\over{M}}{\langle{r}^2\rangle} =\beta + \Delta\beta. \eqno(2)   $$

Here $\vec{d} = \sum{e_k}\vec{r_k}$ and $\vec{\mu}$ are the electric and
magnetic dipole operator respectively. We obtained $[11]$ k = 0.68, 0.60,
0.26, and we use  rms charge radii from lattice QCD calculations $[11,16]$ of
R = 0.86, 0.96, 0.76 fm, for the proton, $\Sigma^+$, and $\Sigma^-$
respectively. The first and second terms in these equations give the intrinsic
and center of charge oscillation contributions, respectively. The sums are
over all E1 and M1 excitations. The intrinsic polarizabilities probe the
internal structure of baryons and mesons.

Many theoretical and experimental polarizability  studies have been made for
the proton and neutron $[17,18]$. In the proton polarizability  calculation of
Weiner and Weise $[18]$, the intrinsic part is mainly due to the charged pion
or kaon cloud surrounding the proton core. Only two calculations were reported
for the $\Sigma$. One is a simple non-relativistic quark bag  calculation
$[4,11]$ with no cloud. The second is via heavy baryon chiral perturbation
theory $[9]$. Calculations are in progress by Lvov $[5]$ for the Sigma Compton
scattering with a dispersion relationship approach. Such an approach is needed
to account for the effects of inelastic reactions such as $\gamma \Sigma
\rightarrow \pi \Lambda$.

The odd parity $\Sigma^*$1/2$^-$ and $\Sigma^*$3/2$^-$ states near 520 MeV
excitation  above the nucleon have the s orbit strange quark promoted to the  p
orbit; and also excitations of the nonstrange quarks. The intrinsic
contribution of eqn. 1 can be evaluated using closure by saturating the sum
over these states, giving $[2,11]$:
$$ {{\alpha}_{\Sigma^+}}\sim{{2}\over{{\Delta}E}}{\langle}\Sigma^+|({\sum}
{q_i}{z_i})^2|\Sigma^+ \rangle^2
{\sim}{{{2}\over{3}}{{\langle{r^2}\rangle}\over {{\Delta}E}}e^2
({{4}\over{9}}+{{4}\over{9}}+{{1}\over{9}}}) \approx 17.1. \eqno(3)   $$

The intrinsic  magnetic polarizability can be evaluated by saturating the
magnetic dipole excitations with the $\Sigma^+$ to $\Sigma(1385)$  transition,
giving $[4,11]$:
$$ \beta_{\Sigma^+}\sim{{2}\over{M_{\Sigma^{*+}}-M_{\Sigma}}}
|{\langle}\Sigma^+|\vec{{\mu}_z}|{\Sigma^*}\rangle|^2
\sim{{2}\over{M_{\Sigma^{*+}}-M_{\Sigma}}}({{2\sqrt{2}}\over{3}}
{\mu}_{\Sigma^+})^2 \sim 8.5, \eqno(4) $$
where ${\mu}_{\Sigma^+}$ is the $\Sigma^+$ magnetic moment. The $\Sigma^+$ to
$\Sigma^*$ magnetic dipole transition matrix element is written in
terms of the $\Sigma^+$ magnetic moment, following the non-relativistic quark
model. For $\beta_{\Sigma^+}$, the magnetic dipole transition to the
$\Sigma^*$ 3/2$^+$ resonance saturates the magnetic dipole excitations, in
analogy to the nucleon to $\Delta$ transition; the proton and $\Sigma^+$ have
the same matrix element expression as members of the same U-spin doublet.  For
the proton intrinsic magnetic polarizability $\beta_p$, the analog of eqn. 4
gives $[4]$ $\beta_p \approx$ 7.4 .

   Consider matrix elements for magnetic moment operators $\vec{\mu}$ between
states having the quark constituents of nucleons and $\Sigma$'s. In the SU(3)
symmetry limit, the contributions to the magnetic moment of the odd $d$ quark
to $\mu_p$ and of the odd $s$ quark to $\mu_{\Sigma}$ are equal. Consider
experimental tests of SU(3) breaking mechanisms based on the assumptions that
the $u$-flavor contributions to $\mu_p$ and $\mu_{\Sigma^+}$ are equal, and
that $SU(3)$ breaking only occurs in $\vec{\mu}_s$. One can then show $[11]$
that the strange contribution to the magnetic moment is suppressed by an order
of magnitude:
$$ {{\mu_{\Sigma^{+}} + 2\mu_{\Sigma^{-}}}\over {\mu_{p} + 2\mu_{n}}} \approx
{{\bra {\Sigma^{+}} \mu_s \ket{\Sigma^{+}}}\over  {\bra {p} \mu_d\ket{p}}}
\approx 0.11 \pm 0.04. \eqno(5) $$
The main physical difficulty is in the small value of the $d$ quark
contribution to the proton moment and the large contribution $2.45$ n.m. of the
$u$ quarks. If the $u$ quark contribution to $\mu_{\Sigma^{+}} $ is the same
$2.45$ n.m. as in the proton, then the experimental value $\mu_{\Sigma^{+}}=
2.42 \pm 0.02$ n.m. can be fit only by requiring a nearly vanishing $s$ quark
contribution.
That the strange quark contribution to $\mu_{\Sigma}$ is suppressed can also be
seen by comparing the strange contributions to $\mu_{\Sigma}$ and
$\mu_{\Lambda}$. The contribution to $\mu_{\Sigma}$ is
only $6\%$ of the strange contribution to $\mu_{\Lambda}$ $[11]$.

Experimental data on related processes may
give additional insight. One such process involves octet-decuplet transitions
for nucleons and $\Sigma$'s. It will be interesting to measure these
transitions and compare their systematics with those of the magnetic moments.
In a description of SU(3) symmetry breaking in the $\Sigma^-$  to
$\Sigma^{*-}$ transition, the ratio of M1 decay widths or of intrinsic
magnetic polarizabilities for the $\Sigma^-$ and proton can be estimated
as $[10-12]$:
$$ {{\Gamma(\Sigma \rightarrow \Sigma^{*-})}\over {\Gamma(p \rightarrow
\Delta^+)}} = {{(M_{\Sigma^{*-}}-M_{\Sigma^-})\beta_{\Sigma^-}}\over
{(M_{\Delta^{+}}-M_{p})\beta_{p}}} \approx {{1}\over{9}}
(1-{{\mu_s}\over{\mu_d}})^2 \approx 0.011.   \eqno(6)   $$
Here $\mu_s$ and $\mu_d$ are empirical magnetic moments of s and d quarks
(estimated in a simple additive quark model $[19]$ using proton, neutron, and
Lambda magnetic moments, as $\mu_u = 1.852\mu_N$, $\mu_d = -0.972\mu_N$,
$\mu_s = -0.613 \mu_N$), the s and d values unequal due to SU(3) symmetry
breaking. As shown in eqn. 5, $\mu_s$ may be significantly lower than the
value fixed by the Lambda magnetic moment; which would significantly increase
the predicted decay width of eqn. 6. Following the assumptions underlying eqns.
5-6, the $\Sigma^-$ M1 transition width directly determines the relative
magnitudes of the s and d-quark magnetic moments. Explicit quark model
calculations $[20]$ give a larger value of 0.034 for the ratio of eqn. 6, with
$\Gamma(\Sigma^{*+})$ = 104 keV $[20]$ and 117 keV $[21]$; and
$\Gamma(\Sigma^{*-})$ = 2.5 keV $[20]$. These calculations correspond to
$\Gamma(\Delta^{+})$ = 74.3 keV using eqn. 15 of Ref. 11. The $\Delta$
radiative width of the model is smaller than the experimental value, but this
should not affect the reliability of the ratio calculation of eqn. 6. One
expects from eqns. (1-4,6) to observe very large and  very small values of the
intrinsic  magnetic polarizability for the $\Sigma^+$ and $\Sigma^-$,
respectively; and similarly for the $\Sigma^*$ radiative decay width. The
predictions $[11]$ $\overline{{\alpha}_{\Sigma^+}}$ =20.8,
$\overline{{\alpha}_{\Sigma^-}}$ =5.7, $\beta_{\Sigma^+}$ = 8.5,
$\beta_{\Sigma^-}$ = 0.12, $\overline{{\beta}_{\Sigma^+}}$ = 1.7 ,
$\overline{{\beta}_{\Sigma^-}}$ = -1.7 satisfy this expectation. Bernard et
al. $[9]$ predict quite different values.

The $\Sigma$ beam at FNAL is polarized, so that asymmetry measurements are
possible; as in the recent study $[22]$ of the $\Sigma \rightarrow p \gamma$
weak decay. Asymmetry data can also be of value for the $\Sigma$
polarizability determination. For radiative transitions, they are sensitive
to the M1 or E2 nature of the exchanged photon, and therefore to the L=2
admixtures. The ratio E2/M1 of these amplitudes  is a subject of considerable
current interest $[23,24]$ for the nucleon to $\Delta$ transition; so that
measuring  this ratio for the $\Sigma^*$ (and $\Xi^*$) will be extremely
valuable. It was suggested $[2,11]$ that the ratios may be different for
$\Sigma^*$, $\Xi^*$, and $\Delta$, given that the s-quark mass is
significantly heavier than the d quark mass. Recent calculations $[25,26]$
give predictions for such transitions.

    We consider now the event rate expectations and backgrounds for the
$\Sigma$ polarizability and radiative transition experiments. We give the
signal and background rates for a C$^{12}$ target. For higher Z targets, the
Z$^2$ dependence of the Primakoff cross section compared to the A$^{2/3}$
dependence of the strong cross section backgrounds, will improve the situation.
We take a t-interval up to 6.0$\times 10^{-3}~GeV^2$, where t is the four
momentum transfer to the target nucleus, as discussed in section 3. We compare
the $\Sigma$ expected rates to data and calculations for the $\pi \rightarrow
\rho \gamma$ radiative transition measurement $[27]$ at 156 GeV on a C$^{12}$
target. For the $\rho$ to $\pi$ transition, for this t-interval, the Primakoff
cross section was 2.4 $\mu$b. and the strong background cross section was 0.75
$\mu$b; which gives a signal to background ratio of 3. The hadronic cross
section falls $[12,28]$ as 1/E (factor=1/3.8), while the Primakoff cross
section rises as ln(E) (factor = 1.3). In addition, the Primakoff cross
section has a coefficient K, given by:
$$ K= {{2J_a^*+1}\over{2J_b^*+1}} ({{{M_a^*}\over{M_{a*}^2-M_a^2}}})^3
\Gamma_{a^*\rightarrow a\gamma}.   \eqno(7) $$
\noindent
Considering mass and spin values, the coefficient K is 7 times larger for the
$\Sigma^* \rightarrow \Sigma$ transition, compared to the $\rho \rightarrow
\pi$ transition. We take $[11]$ also the widths $\Gamma(\Sigma^+)$ = 1000 KeV,
and $\Gamma(\Sigma^-)$= 25 KeV. These are fixed by the widths given above,
normalized to the experimental width of the $\Delta \rightarrow$ N radiative
transition. The theoretical $\Delta$ width calculation is roughly 10 times
lower than the experimental value. We assume here that the model calculation
of width ratios is more accurate than the absolute values. We also assume that
the strong backgrounds for the $\rho$ transition are of the same order as
those of the $\Sigma$ transition. We obtain therefore the rough estimate of
signal cross sections of 314 microbarns ($\Sigma^+$), 7.8 microbarns
($\Sigma^-$), with a background cross section of 0.2 microbarns. The count
rates expected are very good, via comparison to the rates explicitly given
later for pion polarizability, with a  roughly 4.0 microbarn cross section.
Furthermore, the width of the mass distribution for diffractive production of
$\Sigma \rightarrow \Lambda \pi^-$ is roughly 100 MeV, as shown in previous
data, and calculations $[29-31]$ for this process at BNL. Since the $\Sigma^*$
width is 36 MeV, and our mass resolution is 10 MeV, we will gain another
factor of 2.5 from a mass cut.

We consider also the background from the small $\Xi^-$ component in the beam.
This decays 100\% to $\Lambda\pi^-$, and is therefore a background for the
$\Sigma^{*-}$ decay. The mass difference of 64 MeV between $\Xi^-$ and
$\Sigma^{*-}$ is large compared to the expected mass resolution of 10 MeV.
This background can therefore be distinguished in the off-line analysis. The
on-line trigger may reduce this background, by requiring the vertex of the
detected pion to be centered at the target position. Other background
reactions have been considered, but will not be discussed here in detail.
These include the high cross section process of $\Sigma^- \rightarrow \Sigma^-
\pi^0$ diffractive dissociation; and the $\Sigma^- \rightarrow n \pi^-$ decay
of beam particles.

\titlea{3}{Pion Polarizabilities and Radiative Transitions}
For the $\gamma$-$\pi$ interaction at low energy, chiral perturbation theory
($\chi$PT) provides a rigorous way to make predictions; because it stems
directly from QCD and relies only on the solid assumptions of spontaneously
broken SU(3)$_L$ $\times$ SU(3)$_R$ chiral symmetry, Lorentz invariance and low
momentum transfer. Unitarity is achieved by adding pion loop corrections to
lowest order, and the resulting infinite divergences are absorbed into physical
(renormalized) coupling constants L$^r_i$ (tree-level coefficients in
L$^{(4)}$, see Refs. ($[32,33]$). With a perturbative expansion of the
effective Lagrangian limited to terms quartic in the momenta and masses
(O(p$^4$)), the method establishes relationships between different processes in
terms of the L$^r_i$. For example, the radiative pion beta decay and electric
pion polarizability are expressed as: $[7,8,32,33]$:
$$ h_A/h_V = 32\pi^2(L^r_9+L^r_{10}); \bar{\alpha}_{\pi} =
{{4\alpha_f}\over{m_{\pi}f^{2}_{\pi}}}(L^r_9+L^r_{10}); \eqno(8) $$
where f$_\pi$ = 92.4  MeV $[34]$ is the pion decay constant, h$_A$ and h$_V$
are the axial vector and vector coupling constants in the decay, and
$\alpha_f$ is the fine structure constant. The experimental ratio h$_A$/h$_V$
= 0.45 $\pm$ 0.06, leads to $\bar{\alpha}_{\pi}$ = -$\bar{\beta}_{\pi}$ = 2.7
$\pm$ 0.4, where the error shown is due to the uncertainty in the h$_A$/h$_V$
measurement.

Other pion polarizability calculations $[35-37]$ find values for
$\bar{\alpha}_{\pi}$ ranging from $\sim 4 - 14$. Holstein $[7]$ showed that
meson exchange via a pole diagram involving the a$_1$(1260)  resonance provides
the main contribution ($\bar{\alpha}_{\pi}$ = 2.6) to the polarizability. The
E781 high energy pion experiments can obtain new high statistics data for
radiative transitions leading from the pion to the a$_1$(1260), and to other
meson resonances. For a$_1$(1260) $\rightarrow \pi \gamma$, the experimental
width $[38]$ is $\Gamma = 0.64 \pm 0.25$ MeV. Xiong, Shuryak, and Brown (XSB)
$[39]$ estimate this  radiative width to be $\Gamma$ = 1.4 MeV, more than two
times higher than the experimental value $[38]$. With this estimated width,
they calculate the pion polarizability to be $\bar{\alpha}_{\pi}$ = 1.8. A
remeasurement of the a$_1(1260)$ width and of the pion polarizability will
allow checking the consistency of their expected relationship.

For the pion polarizability, Antipov et al. $[40,41]$
measured the $\gamma\pi$ scattering
with 40 GeV pions at Serpukhov via radiative pion scattering in the nuclear
Coulomb field ($\pi^{-}$  +  Z $\rightarrow$  $\pi^{-}$  + Z  + gamma). The
final state gamma ray and pion were detected in coincidence. This
reaction corresponded to  $\gamma$ + $\pi^{-}$ $\rightarrow$  $\gamma$ +
$\pi^{-}$ scattering for laboratory gamma-ray energies in the range 60-600 MeV
on a target $\pi^{-}$ at rest.
The data selection criteria at Serpukhov and
E781 requires one photon and one charged particle in the final state, their
total energy consistent with the beam energy, small t, and other position,
angle, and energy conditions.
Only the angular distribution in the backward
hemisphere were measured at 40 GeV.
The angular acceptance and detector threshold of the
$\gamma$-ray calorimeter in E781 allows measuring more complete angular
distributions than was achievable in the 40 GeV experiment.

The pion electric polarizability $\bar{\alpha_{\pi}}$ was deduced $[40,41]$ in
this low statistics experiment ($\sim$ 7000 events) to be
$ \bar{\alpha}_{\pi} = -\bar{\beta}_{\pi} = 6.8 \pm {1.4}_{stat} \pm
{1.2}_{syst}$,
where it was assumed in the analysis that $\bar{{\alpha}_\pi} +
\bar{{\beta}_\pi} = 0$, as expected theoretically $[7]$. An important result
of s-channel (${\gamma} + {\pi} \rightarrow {\gamma} + {\pi}$) dispersion sum
rules $[4]$ for charged pions is: ${\bar{\alpha}_{\pi}} + {\bar{\beta}_{\pi}}
= 0.39 \pm 0.04.$ This result can be used for high statistics data as a
constraint in the data analysis. Charged pion polarizabilities were also
deduced $[42]$ from $\gamma\gamma\rightarrow\pi^+\pi^-$ data which is related
to the Compton scattering by crossing symmetry. Pennington  $[43]$ claims that
such determinations are very insensitive to the polarizability value. This is
so, since two-photon cross section data at low $\pi\pi$ invariant mass agree
well with calculations for a large range of choices of the (undetermined)
position of the chiral zero. However, very different values of polarizability
are associated with the different choices.

In the radiative pion scattering experiments, it was shown experimentally
$[40,41]$ and theoretically $[44]$ that the Coulomb cross section scales as
Z$^2$ and yields sharp peaks in t-distributions at very small four momentum
transfers to the target nucleus $t \leq 6 \times 10^{-4} (GeV/c)^{2}$.
Background from other processes could easily be estimated and subtracted by
extrapolating in t from events in the region of flat t-distribution of 3-8
$\times 10^{-3} (GeV/c)^{2}$. The sources of these backgrounds are  the
coherent process of pion elastic scattering accompanied by gamma emission
$[44]$, contributions of pion (or rho) rescattering $[42,45]$, and higher cross
section inelastic channels $[27,46]$. Available calculations and data show
that these
backgrounds are manageable, and that the signal to background improves
$[12,28,44]$ with increasing incident energy.

 One must also evaluate electromagnetic corrections to radiative pion
scattering, where the requirement is to measure only single photon
bremsstrahlung emission. Here the detailed properties of the gamma detector
are important, such as the photon detector threshold, t-resolution, and the
two-photon angular resolution. Such calculations were carried out $[47,48]$
for the conditions of the planned 600 GeV FNAL experiment. The
corrections are at
the level of 4\%, and can be easily made.

The polarizability Compton scattering and
chiral anomaly processes in E781 represents a difficult
triggering challenge. They have one negative charged track in the final state,
and one or two $\gamma$-rays. This trigger rate
must match that of the charm component of
the experiment. The trigger problems are deadtime problems generated in the
initial trigger levels when information from the $\gamma$-ray detector is not
available yet for trigger rate suppresssion.  The
first level of hardware trigger T$_0$ is meant to identify a beam particle and
an interaction in the target. However, a T$_0$ trigger with multiplicity 1
will generate a trigger for every beam particle, in contrast to
other trigger types which will fire at the beam interaction rate. To
differentiate between a single negatively-charged Primakoff interaction
product coming from the target and a non-interacting beam track, it is planned
to have three hodoscopes to measure a vertical deflection angle of the
outgoing track after the Primakoff target.
We need to form a coincidence between the projective
elements of the first two hodoscopes to form the track road assuming a beam
particle, and then veto in a window around this road in the third hodoscope.
The required electronics
must give the decision in about 140 nsec, in order to enter the
next trigger level on time.
A simpler beam prescale trigger will also be used; which will help control
systematic uncertainties.

      To specifically illustrate some of the kinematics germane to a FNAL
experiment, the reaction:
$$ {\pi} + Z \rightarrow {\pi}' + Z' + {\gamma}' \eqno(9) $$
is considered for a 600 GeV incident pion, where Z is the nuclear charge. The
4-momentum of each particle is $P_{\pi}$, $P_Z$, $P_{{\pi}'}$, $P_{Z'}$, $k'$,
respectively. In the one photon exchange domain, eqn. 9 is equivalent to:
$$ \gamma + {\pi}  \rightarrow  {\gamma}' + {\pi}', \eqno(10) $$
and the 4-momentum of the incident virtual photon is k = $P_Z$-$P_{Z'}$. The
cross section for the reaction of eqn. 9 is described by the the well tested
Primakoff formalism $[27,49]$ that relates processes involving real photon
interactions to production cross sections involving the exchange of virtual
photons. We have:
$$ { {d{\sigma}} \over {dtdsd{\Omega}} } = { { {Z^2} {\alpha}_f}  \over {\pi} }
{{{|F(t)|}^2}\over{s-{m_{\pi}}^2}} {{t-t_0}\over{t^2}} {
{d{\sigma}_{{\gamma}{\pi}}}\over{d{\Omega}} }, \eqno(11) $$
where  d${\sigma}_{{\gamma}{\pi}}$/d${\Omega}$  is the unpolarized differential
cross section for eqn. 10 (for real photons), t is the square of the
four-momentum transfer to the nucleus, F(t) is the nuclear form factor
(essentially unity at small t-values), $\sqrt{s}$ is the mass of the
$\gamma\pi$ final state, and $t_0$ is the minimum value of t to produce a mass
$\sqrt{s}$. The analysis to
determine polarizabilities by fitting the data use  the known $[40-42,50]$
formula for ${d{\sigma}_{\gamma\pi}}/{d{\Omega}}$. This cross section depends
on the polarizabilities and on s and on t$_1$, the square of the 4-momentum
transfer between initial and final state $\gamma$'s. We have:
$$             t = k^2  \equiv  -M(V)^2 ,\eqno(12) $$
where k is the 4-momentum transferred to the nucleus, and M(V) is the virtual
photon mass. Since $t=2M_{Z}(E_{Z',lab} - M_Z) > 0$, the virtual photon mass is
imaginary. To approximate real pion Compton scattering, the virtual photon must
be almost real; M(V)$<$0.077 GeV corresponding to $t < 6.0 \times 10^{-3}
(GeV/c)^2$ can be required in E781. In addition,
$$          s = (P_{{\pi}'}  + k')^2 \equiv  M(\gamma\pi)^2 ,\eqno(13) $$
where $M(\gamma\pi)$ is the $\gamma\pi$ invariant mass. The minimum value for
t $[51]$ is given by:
$$ t_0 \sim (s-m_{\pi}^2)^2/4|\vec{P_{\pi}}|^2, \eqno(14) $$
corresponding to $t_0 \sim 5.4 \times 10^{-8} (GeV/c)^2$ for
$\sqrt{s}=1.75m_{\pi}$ at 600 GeV incident energy. The maximum of the
differential cross section for reaction (11) occurs at t=2t$_0$, and the
integral to 100t$_0$ gives essentially the entire cross section.

With lead glass detectors $[40,41]$, a t-resolution ${\sigma}_t \sim 6.0
\times 10^{-4} (GeV/c)^2$ was achieved at 40 GeV. The t-resolution sets the
experimental maximum in t for accepted events. The SELEX t-resolution will be
10 times worse. The strong backgrounds are associated with particle exchange,
and such cross sections are known $[12,27,28,44]$ to fall rapidly with
increasing energy. From this point of view, the t-resolution is therefore not
a critical parameter for 600 GeV experiments. The integrated Compton cross
section up to $t = 6.0 \times 10^{-3}$ (GeV/c)$^2$ grows $[44,51]$ as
ln($\vec{P_{\pi}}$), where $\vec{P_{\pi}}$ is the laboratory incident pion
3-momentum. With strong backgrounds that decrease roughly as 1/E, the percent
strong background would decrease for a C$^{12}$ target, from the estimated
$[44]$ 2.5\% at 40 GeV to roughly 0.6\% at 600 GeV at FNAL.

 The energy of the incident virtual photon in the pion rest frame is:
$$ E(V) = (s - {m_{\pi}}^2   + t)/2m_{\pi} \sim  (s - {m_{\pi}}^2)/2m_{\pi}
\eqno(15) $$
at small t; so that the energy of the photon is determined by s. The
elemental cross section at low E(V) is a function of E(V), $cos({\theta})$,
$\bar{{\alpha}}$, $\bar{{\beta}}$; where ${\theta}$ is the Compton scattering
angle in the pion rest frame. In this frame, due to Lorentz contraction, the
nucleus Z represents a transverse virtual photon pulse sweeping past the pion.
One should require that E(V) be sufficiently low in energy, such that $\rho$
meson production does not occur on-shell for an incident $\gamma$-ray on a
target pion at rest. The $\gamma$-ray energy required to produce a $\rho$
meson, not counting the $\rho$ width, is given by
$ \omega  = E(V) = (M_\rho^2-m_\pi^2)/(2m_\pi) = 2.0~GeV. $
Considering the $\rho$ width, one could limit the incident energy to be lower
than 1 GeV, corresponding to s $< 15.3 m_\pi^2$. We will analyze the lower
energy data for polarizability purposes, and at higher energies to understand
the $\rho$ and a$_1$ meson contributions.

Consider the case of 600 GeV incident laboratory pions. The laboratory
outgoing $\gamma$-rays emitted up to 5 mrad, and the corresponding outgoing
pions emitted up to 0.3 mrad. The angular resolution for the pion is roughly
0.04 mrad due to multiple scattering in the Primakoff target and in the
in-beam tracking detectors. The gamma ray energies range from 0 - 400 GeV, and
the corresponding outgoing pion energies range from 200 - 600 GeV. The
corresponding Compton scattering angular range is 0 - 180 degrees in the
${\pi}$ rest frame. In practice, the most forward Compton scattering angles are
less accessible, as they correspond to the larger $\gamma$-ray angles in the
laboratory frame which can miss the $\gamma$-ray detector, and where the
$\gamma$-ray energies are also possibly below the detector threshold. But these
forward angles are anyhow not sensitive to the polarizabilities, as discussed
below.

  We consider the uncertainties achievable for the polarizabilities in the
FNAL E781 experiment, based on Monte Carlo simulations. An important
consideration is the information content $[3,51]$ of the data versus x and s,
where x=cos($\theta$). Consider the fraction of the $\gamma\pi$ Compton cross
section arising from the $\bar{\alpha}$ terms in the $\gamma\pi$ center of
mass cross section expression, for $\bar{\alpha}$=6.8. High s values and back
angles (large t$_1$) have maximal information content for the
polarizabilities. For example, the fraction of the cross section at back
angles due to the polarizability term is only 5\% at E(V) = 140 MeV, and
roughly 30\% at E(V) = 600 MeV.

We consider initially a beam energy of 600 GeV, a C$^{12}$ target, and an
s-range of 2. - 10. m$_{\pi}^2$, corresponding to a Primakoff cross section of
$4.0\mu$b, and E(V) = 70 - 628 MeV. We take the $\pi$-Carbon total cross
section at 600 GeV to be 192. mb, eight times the total inelastic
$\pi$-nucleon cross section at 600 GeV. We assume the simultaneous use of a
5\% Carbon interaction target and also a 0.3\% Pb interaction target. We
calculate below the event rate from the Carbon target, and will in addition
have a factor of roughly 2 times higher rate from the Pb target. The
probability P per inelastic interaction for a Primakoff interaction is then $
P= 4.\times10^{-6}/192.\times10^{-3} = 2.1\times 10^{-5}. $ The number of
interactions planned $[1]$ for E781 is roughly 3.$\times$ 10$^{10}$,
corresponding to roughly 6.3$\times$10$^5$ events for
$4.0\mu$b. Including the Pb target
events, the statistics are roughly 3 times higher. The trigger requirement
for a vertical deflection angle of the pion of more than 100 $\mu$rad cuts the
statistics roughly in half. The losses are mainly events at small s and t$_1$,
which minimally affects polarizability uncertainties. Other
experimental efficiencies and acceptances will lower the statistics.

We cite here some Monte Carlo results for 600 GeV Primakoff experiments on
Carbon, assuming the dispersion sum rule result $\bar{\alpha}$+$\bar{\beta}$ =
0.4 and also $\bar{\alpha}$=6.8. At 600 GeV, including the Pb target, for
580,000 events in an s interval (2.0 - 10.0)m$_\pi^2$, we find by fitting
simulated data that $\bar{\alpha}$ = 7.1 $\pm$ 0.4,
$\bar{\alpha}$+$\bar{\beta}$ = 0.3 $\pm$ 0.1. The overall statistical error
for polarizabilities will be improved from the
expected higher statistics; and by additional data in the
sensitive s-interval 10.-15.3 m$_{\pi}^2$. Data in different s-intervals can
be analyzed to give independent values for the polarizabilities, which will
help control the systematic uncertainties. These Monte Carlo simulations show
that the objective of obtaining pion polarizabilities with significantly
smaller statistical and systematic uncertainties is realistic.

\titlea{4}{Conclusions}
The beams at FNAL and CERN
invite hadron Compton scattering and radiative transition
studies for different particle types, such as $\pi^{+,-}$, $K^{+,-}$, p, $\bar
{p}$, $\Sigma$, $\Xi$, $\Lambda$ hyperons, and others. The E781 experiment was
described. Because these transitions involve the well understood
electromagnetic  process, they provide precision tests of the  quark wave
functions characterizing the configurations of the low lying excited states of
hadrons. The 600 GeV beam energy at FNAL is important to get a good yield for
low t events in the radiative scattering, and also to reduce backgrounds from
the decay of unstable hadrons by significantly boosting their lifetime.    We
will measure the $\gamma\pi$ and $\gamma\Sigma$ Compton scattering cross
sections, thereby enabling determinations of the pion and Sigma
polarizabilities. E781 will also measure the formation and decay of the
$\Sigma^-$(1385), $\Sigma^+$(1385), and a$_1$(1260) excited states. These
various $\Sigma$ and pion experiments will allow serious tests of  chiral
symmetry and chiral perturbation theory; and of different available
polarizability and radiative decay calculations in QCD. The $\Sigma$
experiments will shed new light on puzzles associated with the size of the
s-quark magnetic moment.

\titlea{5}{Acknowledgements}
This work was supported by the U.S.-Israel Binational Science Foundation,
Jerusalem, Israel. Thanks are due to  S. Bellucci, P. Cooper, T. Ferbel, L.
Frankfort, S. Gerzon, G. Giordano, A. Kulyavtsev,
J. Lach, H. J. Lipkin, A. Lvov, J. Russ, and N. Terentyev
for valuable discussions.

\begrefchapter{References}
\refno {1.} J. Russ, spokesman, FNAL E781 Collaboration: Carnegie-Mellon U.,
Fermilab, U. Iowa, U. Rochester, U. Washington, Petersburg Nuclear Physics
Institute, ITEP (Moscow), IHEP (Protvino), Moscow State U., U. Sao Paulo,
Centro Brasileiro de Pesquisas Fisicas, Universidade Federale de Paraiba, IHEP
(Beijing), U. Bristol, Tel Aviv U., Max Planck Institut fur
Kernphysik-Heidelberg, Universidad Autonoma de San Luis Potosi; \newline
J. Russ: Proceedings of the CHARM2000 Workshop, Fermilab, June 1994, Eds. D.
M. Kaplan and S. Kwan, Fermilab-Conf-94/190, P. 111, (1994) \refno {2.} M. A.
Moinester: Proceedings of the Conference on the Intersections Between Particle
and Nuclear Physics, Tucson, Arizona, 1991, AIP Conference Proceedings 243, P.
553, 1992, Ed. W. Van Oers.
\refno {3.} M.A. Moinester: Workshop on Hadron Structure from Photo-Reactions
at Intermediate Energies, Brookhaven National Laboratory, May 1992, Eds. A.
Nathan, A. Sandorfi, BNL Report 47972, Tel Aviv U. preprint TAUP 1972/92.
\refno {4.} V. A. Petrunkin: Sov. J. Part. Nucl. {\bf12} 278 (1981)
\refno {5.} A. I. Lvov: Sov. J. Nucl. Phys. {\bf42} 583 (1985);
Int. J. Mod. Phys. A {\bf8}  5267 (1993);
Phys. Lett. B {\bf304} 29 (1993) ; private communication.
\refno {6.} J. L. Friar: Workshop on Electron-Nucleus Scattering (1988, EIPC),
 Eds. A. Fabrocini  et al., World Scientific Publishing Co. (1989)
\refno {7.}   B. R. Holstein: Comments Nucl. Part. Phys. {\bf19}  239 (1990)
\refno {8.} J. F. Donoghue, B. R. Holstein: Phys. Rev. D {\bf40} 2378 (1989)
\refno {9.} V. Bernard et al: Phys. Rev. D {\bf46}  2756 (1992); Phys. Lett.
B {\bf319} 269 (1993)
\refno {10.} H. J. Lipkin: Phys. Rev. D {\bf7} 846 (1973)
\refno {11.} H. J. Lipkin, M. A. Moinester: Phys. Lett. B {\bf287} 179 (1992)
\refno {12.} A. V. Vanyashin et al: Sov. J. Nucl. Phys. {\bf34} 90 (1981)
\refno {13.} T. Goldman  et al: Physics with LAMPF II, LA-9798-P, P. 319 (1984)
\refno {14.} M. V. Hynes: Physics with LAMPF II, LA-9798-P, P. 333 (1984)
\refno {15.} F. Dydak et al: Nucl. Phys. B {\bf118} 1 (1977)
\refno {16.} D. B. Leinweber, R. M. Woloshyn, T. Draper: Phys. Rev. D {\bf43}
1659 (1991) ; \newline
D. B. Leinweber, T. D. Cohen: Phys. Rev. D {\bf47} 2147 (1993)
\refno {17.} B. R. Holstein, A. M. Nathan: Phys. Rev. D {\bf49} 6101 (1994)
\refno {18.} R. Weiner, W. Weise: Phys. Lett. B {\bf159} 85 (1985)
\refno {19.} Particle Data Group, G.P.Yost et al: Phys. Lett. B {\bf204} 1
(1988)
\refno {20.} J. W. Darewych  et al: Phys. Rev. D {\bf28} 1125 (1983)
\refno {21.} E. Kaxiras, E. J. Moniz, M. Soyeur:
Phys. Rev. D {\bf32} 695, (1985)
\refno {22.} M. Foucher et al: Phys. Rev. Lett. {\bf68}  3004 (1992)
\refno {23.} R. M. Davidson, N. C. Mukhopadhyay, R.S. Wittman: Phys. Rev. D
{\bf43} 71 (1991)
\refno {24.} A. Bernstein, S. Nozawa, M. A. Moinester: Phys. Rev. C {\bf47}
1274 (1993)
\refno {25.} D. B. Leinweber, T. Draper, R. M. Woloshyn: Phys. Rev. D {\bf48}
2230 (1993)
\refno {26.} M. N. Butler, M. J. Savage, R. P. Springer: Phys. Lett. B
{\bf304} 353 (1993); \newline
Phys. Lett. {\bf314} 122(E)(1993); Nucl. Phys. B {\bf399} 69 (1993)
\refno {27.} T. Jensen et al: Phys. Rev. D {\bf27} 26 (1983)
\refno {28.} G. Berlad et al: Ann. Phys. {\bf 75} 461 (1973)
\refno {29.} R. T. Deck: Phys. Rev. Lett. {\bf13} 169 (1964)
\refno {30.} L. Stodolsky: Phys. Rev. Lett. {\bf18} 973 (1967)
\refno {31.} V. Hungerbuhler et al: Phys. Rev. D {\bf10} 2051 (1974)
\refno {32.} J.Gasser, H.Leutwyler: Nucl. Phys. B {\bf250} 465 (1985)
\refno {33.} J.Gasser, H.Leutwyler: Ann.Phys. (N.Y.) {\bf 158} 142 (1984)
\refno {34.} B.R.Holstein: Phys.Lett.B {\bf 244} 83 (1990)
\refno {35.} V. Bernard, B. Hiller, W. Weise: Phys. Lett.  B {\bf159} 85 (1988)
\refno {36.} V. Bernard, D. Vautherin: Phys. Rev. D {\bf40} 1615 (1989)
\refno {37.} M. A. Ivanov, T. Mizutani: Phys. Rev. D {\bf45} 1580 (1992)
\refno {38.} M. Zielinski et al: Phys. Rev. Lett. {\bf52} 1195 (1984)
\refno {39.} L. Xiong, E. Shuryak, G. Brown: Phys. Rev. D {\bf46} 3798 (1992)
\refno {40.} Y. M. Antipov et al: Phys. Lett. B {\bf121} 445 (1983)
\refno {41.} Y. M. Antipov et al: Z. Phys. C-Particles and Fields {\bf26}
495 (1985)
\refno {42.} D. Babusci, S. Bellucci, G. Giordano, G. Matone,
A. M. Sandorfi, M. A. Moinester: Phys. Lett. B {\bf277} 158 (1992)
\refno {43.} J. Portoles, M. R. Pennington: U. Durham preprint DTP-94/52,
1994, submitted to Second DA$\Phi$NE Physics Handbook, Eds. G. Pancheri and N.
Paver.
\refno {44.} A. S. Galperin et al: Sov. J. Nucl. Phys. {\bf32} 545 (1980)
\refno {45.} G. V. Mitselmakher, V. N. Pervushkin: Sov. J. Nucl. Phys.
{\bf37} (1983)
\refno {46.} M. Zielinski et al: Z. Phys. C {\bf16} 197 (1983)
\refno {47.} A. A. Akhundov, D. Yu. Bardin, G. V. Mitselmakher,
A. G. Olshevsky: Sov. J. Nucl. Phys. {\bf42} 426 (1984)
\refno {48.} A. A. Akhundov, S. Gerzon, S. Kananov, M. A. Moinester:
I.C.T.P. (Trieste) Preprint IC/94/203; Tel Aviv U. Preprint TAUP-2183-94,
1994
\refno {49.} M. Zielinski et al: Phys. Rev. D {\bf29} 2633 (1984)
\refno {50.} L. V. Fil'kov, I. Guiasu, E. E. Radescu: Phys. Rev. {\bf26}
3146 (1982)
\refno {51.} N. I. Starkov, L. V. Fil'kov, V. A. Tsarev: Sov. J. Nucl. Phys.
{\bf36} 707 (1982)
\endref
\byebye

\magnification=\magstep0
\font \authfont               = cmr10 scaled\magstep4
\font \fivesans               = cmss10 at 5pt
\font \headfont               = cmbx12 scaled\magstep4
\font \markfont               = cmr10 scaled\magstep1
\font \ninebf                 = cmbx9
\font \ninei                  = cmmi9
\font \nineit                 = cmti9
\font \ninerm                 = cmr9
\font \ninesans               = cmss10 at 9pt
\font \ninesl                 = cmsl9
\font \ninesy                 = cmsy9
\font \ninett                 = cmtt9
\font \sevensans              = cmss10 at 7pt
\font \sixbf                  = cmbx6
\font \sixi                   = cmmi6
\font \sixrm                  = cmr6
\font \sixsans                = cmss10 at 6pt
\font \sixsy                  = cmsy6
\font \smallescriptfont       = cmr5 at 7pt
\font \smallescriptscriptfont = cmr5
\font \smalletextfont         = cmr5 at 10pt
\font \subhfont               = cmr10 scaled\magstep4
\font \tafonts                = cmbx7  scaled\magstep2
\font \tafontss               = cmbx5  scaled\magstep2
\font \tafontt                = cmbx10 scaled\magstep2
\font \tams                   = cmmib10
\font \tamss                  = cmmib10 scaled 700
\font \tamt                   = cmmib10 scaled\magstep2
\font \tass                   = cmsy7  scaled\magstep2
\font \tasss                  = cmsy5  scaled\magstep2
\font \tast                   = cmsy10 scaled\magstep2
\font \tasys                  = cmex10 scaled\magstep1
\font \tasyt                  = cmex10 scaled\magstep2
\font \tbfonts                = cmbx7  scaled\magstep1
\font \tbfontss               = cmbx5  scaled\magstep1
\font \tbfontt                = cmbx10 scaled\magstep1
\font \tbms                   = cmmib10 scaled 833
\font \tbmss                  = cmmib10 scaled 600
\font \tbmt                   = cmmib10 scaled\magstep1
\font \tbss                   = cmsy7  scaled\magstep1
\font \tbsss                  = cmsy5  scaled\magstep1
\font \tbst                   = cmsy10 scaled\magstep1
\font \tenbfne                = cmb10
\font \tensans                = cmss10
\font \tpfonts                = cmbx7  scaled\magstep3
\font \tpfontss               = cmbx5  scaled\magstep3
\font \tpfontt                = cmbx10 scaled\magstep3
\font \tpmt                   = cmmib10 scaled\magstep3
\font \tpss                   = cmsy7  scaled\magstep3
\font \tpsss                  = cmsy5  scaled\magstep3
\font \tpst                   = cmsy10 scaled\magstep3
\font \tpsyt                  = cmex10 scaled\magstep3
\vsize=19.3cm
\hsize=12.2cm
\hfuzz=2pt
\tolerance=500
\abovedisplayskip=3 mm plus6pt minus 4pt
\belowdisplayskip=3 mm plus6pt minus 4pt
\abovedisplayshortskip=0mm plus6pt minus 2pt
\belowdisplayshortskip=2 mm plus4pt minus 4pt
\predisplaypenalty=0
\clubpenalty=10000
\widowpenalty=10000
\frenchspacing
\newdimen\oldparindent\oldparindent=1.5em
\parindent=1.5em
\skewchar\ninei='177 \skewchar\sixi='177
\skewchar\ninesy='60 \skewchar\sixsy='60
\hyphenchar\ninett=-1
\def\newline{\hfil\break}%
\catcode`@=11
\def\folio{\ifnum\pageno<\z@
\uppercase\expandafter{\romannumeral-\pageno}%
\else\number\pageno \fi}
\catcode`@=12 
  \mathchardef\Gamma="0100
  \mathchardef\Delta="0101
  \mathchardef\Theta="0102
  \mathchardef\Lambda="0103
  \mathchardef\Xi="0104
  \mathchardef\Pi="0105
  \mathchardef\Sigma="0106
  \mathchardef\Upsilon="0107
  \mathchardef\Phi="0108
  \mathchardef\Psi="0109
  \mathchardef\Omega="010A
  \mathchardef\bfGamma="0\the\bffam 00
  \mathchardef\bfDelta="0\the\bffam 01
  \mathchardef\bfTheta="0\the\bffam 02
  \mathchardef\bfLambda="0\the\bffam 03
  \mathchardef\bfXi="0\the\bffam 04
  \mathchardef\bfPi="0\the\bffam 05
  \mathchardef\bfSigma="0\the\bffam 06
  \mathchardef\bfUpsilon="0\the\bffam 07
  \mathchardef\bfPhi="0\the\bffam 08
  \mathchardef\bfPsi="0\the\bffam 09
  \mathchardef\bfOmega="0\the\bffam 0A

\def\sq{\hbox{\rlap{$\sqcap$}$\sqcup$}}

\def\utw{\smash{\rlap{\lower5pt\hbox{$\sim$}}}}
\def\udtw{\smash{\rlap{\lower6pt\hbox{$\approx$}}}}

\def\diameter{{\ifmmode\mathchoice
{\ooalign{\hfil\hbox{$\displaystyle/$}\hfil\crcr
{\hbox{$\displaystyle\mathchar"20D$}}}}
{\ooalign{\hfil\hbox{$\textstyle/$}\hfil\crcr
{\hbox{$\textstyle\mathchar"20D$}}}}
{\ooalign{\hfil\hbox{$\scriptstyle/$}\hfil\crcr
{\hbox{$\scriptstyle\mathchar"20D$}}}}
{\ooalign{\hfil\hbox{$\scriptscriptstyle/$}\hfil\crcr
{\hbox{$\scriptscriptstyle\mathchar"20D$}}}}
\else{\ooalign{\hfil/\hfil\crcr\mathhexbox20D}}%
\fi}}


\def\bbbc{{\mathchoice {\setbox0=\hbox{$\displaystyle\rm C$}\hbox{\hbox
to0pt{\kern0.4\wd0\vrule height0.9\ht0\hss}\box0}}
{\setbox0=\hbox{$\textstyle\rm C$}\hbox{\hbox
to0pt{\kern0.4\wd0\vrule height0.9\ht0\hss}\box0}}
{\setbox0=\hbox{$\scriptstyle\rm C$}\hbox{\hbox
to0pt{\kern0.4\wd0\vrule height0.9\ht0\hss}\box0}}
{\setbox0=\hbox{$\scriptscriptstyle\rm C$}\hbox{\hbox
to0pt{\kern0.4\wd0\vrule height0.9\ht0\hss}\box0}}}}
\def\bbbe{{\mathchoice {\setbox0=\hbox{\smalletextfont e}\hbox{\raise
0.1\ht0\hbox to0pt{\kern0.4\wd0\vrule width0.3pt height0.7\ht0\hss}\box0}}
{\setbox0=\hbox{\smalletextfont e}\hbox{\raise
0.1\ht0\hbox to0pt{\kern0.4\wd0\vrule width0.3pt height0.7\ht0\hss}\box0}}
{\setbox0=\hbox{\smallescriptfont e}\hbox{\raise
0.1\ht0\hbox to0pt{\kern0.5\wd0\vrule width0.2pt height0.7\ht0\hss}\box0}}
{\setbox0=\hbox{\smallescriptscriptfont e}\hbox{\raise
0.1\ht0\hbox to0pt{\kern0.4\wd0\vrule width0.2pt height0.7\ht0\hss}\box0}}}}
\def\bbbq{{\mathchoice {\setbox0=\hbox{$\displaystyle\rm Q$}\hbox{\raise
0.15\ht0\hbox to0pt{\kern0.4\wd0\vrule height0.8\ht0\hss}\box0}}
{\setbox0=\hbox{$\textstyle\rm Q$}\hbox{\raise
0.15\ht0\hbox to0pt{\kern0.4\wd0\vrule height0.8\ht0\hss}\box0}}
{\setbox0=\hbox{$\scriptstyle\rm Q$}\hbox{\raise
0.15\ht0\hbox to0pt{\kern0.4\wd0\vrule height0.7\ht0\hss}\box0}}
{\setbox0=\hbox{$\scriptscriptstyle\rm Q$}\hbox{\raise
0.15\ht0\hbox to0pt{\kern0.4\wd0\vrule height0.7\ht0\hss}\box0}}}}
\def\bbbt{{\mathchoice {\setbox0=\hbox{$\displaystyle\rm
T$}\hbox{\hbox to0pt{\kern0.3\wd0\vrule height0.9\ht0\hss}\box0}}
{\setbox0=\hbox{$\textstyle\rm T$}\hbox{\hbox
to0pt{\kern0.3\wd0\vrule height0.9\ht0\hss}\box0}}
{\setbox0=\hbox{$\scriptstyle\rm T$}\hbox{\hbox
to0pt{\kern0.3\wd0\vrule height0.9\ht0\hss}\box0}}
{\setbox0=\hbox{$\scriptscriptstyle\rm T$}\hbox{\hbox
to0pt{\kern0.3\wd0\vrule height0.9\ht0\hss}\box0}}}}
\def\bbbs{{\mathchoice
{\setbox0=\hbox{$\displaystyle     \rm S$}\hbox{\raise0.5\ht0\hbox
to0pt{\kern0.35\wd0\vrule height0.45\ht0\hss}\hbox
to0pt{\kern0.55\wd0\vrule height0.5\ht0\hss}\box0}}
{\setbox0=\hbox{$\textstyle        \rm S$}\hbox{\raise0.5\ht0\hbox
to0pt{\kern0.35\wd0\vrule height0.45\ht0\hss}\hbox
to0pt{\kern0.55\wd0\vrule height0.5\ht0\hss}\box0}}
{\setbox0=\hbox{$\scriptstyle      \rm S$}\hbox{\raise0.5\ht0\hbox
to0pt{\kern0.35\wd0\vrule height0.45\ht0\hss}\raise0.05\ht0\hbox
to0pt{\kern0.5\wd0\vrule height0.45\ht0\hss}\box0}}
{\setbox0=\hbox{$\scriptscriptstyle\rm S$}\hbox{\raise0.5\ht0\hbox
to0pt{\kern0.4\wd0\vrule height0.45\ht0\hss}\raise0.05\ht0\hbox
to0pt{\kern0.55\wd0\vrule height0.45\ht0\hss}\box0}}}}
\def\bbbz{{\mathchoice {\hbox{$\sans\textstyle Z\kern-0.4em Z$}}
{\hbox{$\sans\textstyle Z\kern-0.4em Z$}}
{\hbox{$\sans\scriptstyle Z\kern-0.3em Z$}}
{\hbox{$\sans\scriptscriptstyle Z\kern-0.2em Z$}}}}
\def\qed{\ifmmode\sq\else{\unskip\nobreak\hfil
\penalty50\hskip1em\null\nobreak\hfil\sq
\parfillskip=0pt\finalhyphendemerits=0\endgraf}\fi}
\newfam\sansfam
\textfont\sansfam=\tensans\scriptfont\sansfam=\sevensans
\scriptscriptfont\sansfam=\fivesans
\def\sans{\fam\sansfam\tensans}
\def\stackfigbox{\if
Y\FIG\global\setbox\figbox=\vbox{\unvbox\figbox\box1}%
\else\global\setbox\figbox=\vbox{\box1}\global\let\FIG=Y\fi}
\def\placefigure{\dimen0=\ht1\advance\dimen0by\dp1
\advance\dimen0by5\baselineskip
\advance\dimen0by0.33333 cm
\ifdim\dimen0>\vsize\pageinsert\box1\vfill\endinsert
\else
\if Y\FIG\stackfigbox\else
\dimen0=\pagetotal\ifdim\dimen0<\pagegoal
\advance\dimen0by\ht1\advance\dimen0by\dp1\advance\dimen0by1.16666cm
\ifdim\dimen0>\pagegoal\stackfigbox
\else\box1\vskip3.33333 mm\fi
\else\box1\vskip3.33333 mm\fi\fi\fi}
%
\def\begfig#1cm#2\endfig{\par
\setbox1=\vbox{\dimen0=#1true cm\advance\dimen0
by0.83333 cm\kern\dimen0#2}\placefigure}
\def\begdoublefig#1cm #2 #3 \enddoublefig{\begfig#1cm%
\vskip-.8333\baselineskip\line{\vtop{\hsize=0.46\hsize#2}\hfill
\vtop{\hsize=0.46\hsize#3}}\endfig}
\def\begfigsidebottom#1cm#2cm#3\endfigsidebottom{\dimen0=#2true cm
\ifdim\dimen0<0.4\hsize\message{Room for legend to narrow;
begfigsidebottom changed to begfig}\begfig#1cm#3\endfig\else
\par\def\figure##1##2{\vbox{\noindent\petit{\bf
Fig.\ts##1\unskip.\ }\ignorespaces ##2\par}}%
\dimen0=\hsize\advance\dimen0 by-.66666 cm\advance\dimen0 by-#2true cm
\setbox1=\vbox{\hbox{\hbox to\dimen0{\vrule height#1true cm\hrulefill}%
\kern.66666 cm\vbox{\hsize=#2true cm#3}}}\placefigure\fi}
\def\begfigsidetop#1cm#2cm#3\endfigsidetop{\dimen0=#2true cm
\ifdim\dimen0<0.4\hsize\message{Room for legend to narrow; begfigsidetop
changed to begfig}\begfig#1cm#3\endfig\else
\par\def\figure##1##2{\vbox{\noindent\petit{\bf
Fig.\ts##1\unskip.\ }\ignorespaces ##2\par}}%
\dimen0=\hsize\advance\dimen0 by-.66666 cm\advance\dimen0 by-#2true cm
\setbox1=\vbox{\hbox{\hbox to\dimen0{\vrule height#1true cm\hrulefill}%
\kern.66666 cm\vbox to#1true cm{\hsize=#2true cm#3\vfill
}}}\placefigure\fi}
\def\figure#1#2{\vskip0.83333 cm\setbox0=\vbox{\noindent\petit{\bf
Fig.\ts#1\unskip.\ }\ignorespaces #2\smallskip
\count255=0\global\advance\count255by\prevgraf}%
\ifnum\count255>1\box0\else
\centerline{\petit{\bf Fig.\ts#1\unskip.\
}\ignorespaces#2}\smallskip\fi}

\def\begtab#1cm#2\endtab{\par
   \ifvoid\topins\midinsert\medskip\vbox{#2\kern#1true cm}\endinsert
   \else\topinsert\vbox{#2\kern#1true cm}\endinsert\fi}
\def\begpet{\vskip6pt\bgroup\petit}
\def\endpet{\vskip6pt\egroup}
\newcount\frpages
\newcount\frpagegoal
\def\freepage#1{\global\frpagegoal=#1\relax\global\frpages=0\relax
\loop\global\advance\frpages by 1\relax
\message{Doing freepage \the\frpages\space of
\the\frpagegoal}\null\vfill\eject
\ifnum\frpagegoal>\frpages\repeat}
\newdimen\refindent
\def\begrefchapter#1{\titlea{}{\ignorespaces#1}%
\bgroup\petit
\setbox0=\hbox{1000.\enspace}\refindent=\wd0}
\def\ref{\goodbreak
\hangindent\oldparindent\hangafter=1
\noindent\ignorespaces}
\def\refno#1{\goodbreak
\hangindent\refindent\hangafter=1
\noindent\hbox to\refindent{#1\hss}\ignorespaces}
\def\endref{\goodbreak\endpet}
\def\vec#1{{\textfont1=\tams\scriptfont1=\tamss
\textfont0=\tenbf\scriptfont0=\sevenbf
\mathchoice{\hbox{$\displaystyle#1$}}{\hbox{$\textstyle#1$}}
{\hbox{$\scriptstyle#1$}}{\hbox{$\scriptscriptstyle#1$}}}}
\def\petit{\def\rm{\fam0\ninerm}%
\textfont0=\ninerm \scriptfont0=\sixrm \scriptscriptfont0=\fiverm
 \textfont1=\ninei \scriptfont1=\sixi \scriptscriptfont1=\fivei
 \textfont2=\ninesy \scriptfont2=\sixsy \scriptscriptfont2=\fivesy
 \def\it{\fam\itfam\nineit}%
 \textfont\itfam=\nineit
 \def\sl{\fam\slfam\ninesl}%
 \textfont\slfam=\ninesl
 \def\bf{\fam\bffam\ninebf}%
 \textfont\bffam=\ninebf \scriptfont\bffam=\sixbf
 \scriptscriptfont\bffam=\fivebf
 \def\sans{\fam\sansfam\ninesans}%
 \textfont\sansfam=\ninesans \scriptfont\sansfam=\sixsans
 \scriptscriptfont\sansfam=\fivesans
 \def\tt{\fam\ttfam\ninett}%
 \textfont\ttfam=\ninett
 \normalbaselineskip=11pt
 \setbox\strutbox=\hbox{\vrule height7pt depth2pt width0pt}%
 \normalbaselines\rm
\def\vec##1{{\textfont1=\tbms\scriptfont1=\tbmss
\textfont0=\ninebf\scriptfont0=\sixbf
\mathchoice{\hbox{$\displaystyle##1$}}{\hbox{$\textstyle##1$}}
{\hbox{$\scriptstyle##1$}}{\hbox{$\scriptscriptstyle##1$}}}}}
\nopagenumbers
%
\let\header=Y
\let\FIG=N
\newbox\figbox
\output={\if N\header\headline={\hfil}\fi\plainoutput\global\let\header=Y
\if Y\FIG\topinsert\unvbox\figbox\endinsert\global\let\FIG=N\fi}
\let\lasttitle=N
\def\bookauthor#1{\vfill\eject
     \bgroup
     \baselineskip=22pt
     \lineskip=0pt
     \pretolerance=10000
     \authfont
     \rightskip 0pt plus 6em
     \centerpar{#1}\vskip1.66666 cm\egroup}
\def\bookhead#1#2{\bgroup
     \baselineskip=36pt
     \lineskip=0pt
     \pretolerance=10000
     \headfont
     \rightskip 0pt plus 6em
     \centerpar{#1}\vskip0.83333 cm
     \baselineskip=22pt
     \subhfont\centerpar{#2}\vfill
     \parindent=0pt
     \baselineskip=16pt
     \leftskip=1.83333cm
     \markfont Springer-Verlag\newline
     Berlin Heidelberg New York\newline
     London Paris Tokyo Singapore\bigskip\bigskip
     [{\it This is page III of your manuscript and will be reset by
     Springer.}]
     \egroup\let\header=N\eject}
\def\centerpar#1{{\parfillskip=0pt
\rightskip=0pt plus 1fil
\leftskip=0pt plus 1fil
\advance\leftskip by\oldparindent
\advance\rightskip by\oldparindent
\def\newline{\break}%
\noindent\ignorespaces#1\par}}
\def\part#1#2{\vfill\supereject\let\header=N
\centerline{\subhfont#1}%
\vskip75pt
     \bgroup
\textfont0=\tpfontt \scriptfont0=\tpfonts \scriptscriptfont0=\tpfontss
\textfont1=\tpmt \scriptfont1=\tbmt \scriptscriptfont1=\tams
\textfont2=\tpst \scriptfont2=\tpss \scriptscriptfont2=\tpsss
\textfont3=\tpsyt \scriptfont3=\tasys \scriptscriptfont3=\tenex
     \baselineskip=20pt
     \lineskip=0pt
     \pretolerance=10000
     \tpfontt
     \centerpar{#2}
     \vfill\eject\egroup\ignorespaces}
\newtoks\AUTHOR
\newtoks\HEAD
\catcode`\@=\active
\def\author#1{\bgroup
\baselineskip=22pt
\lineskip=0pt
\pretolerance=10000
\markfont
\centerpar{#1}\bigskip\egroup
{\def@##1{}%
\setbox0=\hbox{\petit\kern2.08333 cc\ignorespaces#1\unskip}%
\ifdim\wd0>\hsize
\message{The names of the authors exceed the headline, please use a }%
\message{short form with AUTHORRUNNING}\gdef\leftheadline{%
\hbox to2.08333 cc{\folio\hfil}AUTHORS suppressed due to excessive
length\hfil}%
\global\AUTHOR={AUTHORS were to long}\else
\xdef\leftheadline{\hbox to2.08333
cc{\noexpand\folio\hfil}\ignorespaces#1\hfill}%
\global\AUTHOR={\def@##1{}\ignorespaces#1\unskip}\fi
}\let\INS=E}
\def\address#1{\bgroup
\centerpar{#1}\bigskip\egroup
\catcode`\@=12
\vskip2cm\noindent\ignorespaces}
\let\INS=N%
\def@#1{\if N\INS\unskip\ $^{#1}$\else\if
E\INS\noindent$^{#1}$\let\INS=Y\ignorespaces
\else\par
\noindent$^{#1}$\ignorespaces\fi\fi}%
\catcode`\@=12
\headline={\petit\def\newline{ }\def\fonote#1{}\ifodd\pageno
\rightheadline\else\leftheadline\fi}
\def\rightheadline{\hfil Missing CONTRIBUTION
title\hbox to2.08333 cc{\hfil\folio}}
\def\leftheadline{\hbox to2.08333 cc{\folio\hfil}Missing name(s) of the
author(s)\hfil}
\nopagenumbers
\let\header=Y

\let\lasttitle=N
 \def\contribution#1{\vfill\supereject
 \ifodd\pageno\else\null\vfill\supereject\fi
 \let\header=N\bgroup
 \textfont0=\tafontt \scriptfont0=\tafonts \scriptscriptfont0=\tafontss
 \textfont1=\tamt \scriptfont1=\tams \scriptscriptfont1=\tams
 \textfont2=\tast \scriptfont2=\tass \scriptscriptfont2=\tasss
 \par\baselineskip=16pt
     \lineskip=16pt
     \tafontt
     \raggedright
     \pretolerance=10000
     \noindent
     \centerpar{\ignorespaces#1}%
     \vskip12pt\egroup
     \nobreak
     \parindent=0pt
     \everypar={\global\parindent=1.5em
     \global\let\lasttitle=N\global\everypar={}}%
     \global\let\lasttitle=A%
     \setbox0=\hbox{\petit\def\newline{ }\def\fonote##1{}\kern2.08333
     cc\ignorespaces#1}\ifdim\wd0>\hsize
     \message{Your CONTRIBUTIONtitle exceeds the headline,
please use a short form
with CONTRIBUTIONRUNNING}\gdef\rightheadline{\hfil CONTRIBUTION title
suppressed due to excessive length\hbox to2.08333 cc{\hfil\folio}}%
\global\HEAD={HEAD was to long}\else
\gdef\rightheadline{\hfill\ignorespaces#1\unskip\hbox to2.08333
cc{\hfil\folio}}\global\HEAD={\ignorespaces#1\unskip}\fi
\catcode`\@=\active
     \ignorespaces}
 \def\contributionnext#1{\vfill\supereject
 \let\header=N\bgroup
 \textfont0=\tafontt \scriptfont0=\tafonts \scriptscriptfont0=\tafontss
 \textfont1=\tamt \scriptfont1=\tams \scriptscriptfont1=\tams
 \textfont2=\tast \scriptfont2=\tass \scriptscriptfont2=\tasss
 \par\baselineskip=16pt
     \lineskip=16pt
     \tafontt
     \raggedright
     \pretolerance=10000
     \noindent
     \centerpar{\ignorespaces#1}%
     \vskip12pt\egroup
     \nobreak
     \parindent=0pt
     \everypar={\global\parindent=1.5em
     \global\let\lasttitle=N\global\everypar={}}%
     \global\let\lasttitle=A%
     \setbox0=\hbox{\petit\def\newline{ }\def\fonote##1{}\kern2.08333
     cc\ignorespaces#1}\ifdim\wd0>\hsize
     \message{Your CONTRIBUTIONtitle exceeds the headline,
please use a short form
with CONTRIBUTIONRUNNING}\gdef\rightheadline{\hfil CONTRIBUTION title
suppressed due to excessive length\hbox to2.08333 cc{\hfil\folio}}%
\global\HEAD={HEAD was to long}\else
\gdef\rightheadline{\hfill\ignorespaces#1\unskip\hbox to2.08333
cc{\hfil\folio}}\global\HEAD={\ignorespaces#1\unskip}\fi
\catcode`\@=\active
     \ignorespaces}
\def\motto#1#2{\bgroup\petit\leftskip=5.41666cm\noindent\ignorespaces#1
\if!#2!\else\medskip\noindent\it\ignorespaces#2\fi\bigskip\egroup
\let\lasttitle=M
\parindent=0pt
\everypar={\global\parindent=\oldparindent
\global\let\lasttitle=N\global\everypar={}}%
\global\let\lasttitle=M%
\ignorespaces}
\def\abstract#1{\bgroup\petit\noindent
{\bf Abstract: }\ignorespaces#1\vskip28pt\egroup
\let\lasttitle=N
\parindent=0pt
\everypar={\global\parindent=\oldparindent
\global\let\lasttitle=N\global\everypar={}}%
\ignorespaces}
\def\titlea#1#2{\if N\lasttitle\else\vskip-28pt
     \fi
     \vskip18pt plus 4pt minus4pt
     \bgroup
\textfont0=\tbfontt \scriptfont0=\tbfonts \scriptscriptfont0=\tbfontss
\textfont1=\tbmt \scriptfont1=\tbms \scriptscriptfont1=\tbmss
\textfont2=\tbst \scriptfont2=\tbss \scriptscriptfont2=\tbsss
\textfont3=\tasys \scriptfont3=\tenex \scriptscriptfont3=\tenex
     \baselineskip=16pt
     \lineskip=0pt
     \pretolerance=10000
     \noindent
     \tbfontt
     \rightskip 0pt plus 6em
     \setbox0=\vbox{\vskip23pt\def\fonote##1{}%
     \noindent
     \if!#1!\ignorespaces#2
     \else\setbox0=\hbox{\ignorespaces#1\unskip\ }\hangindent=\wd0
     \hangafter=1\box0\ignorespaces#2\fi
     \vskip18pt}%
     \dimen0=\pagetotal\advance\dimen0 by-\pageshrink
     \ifdim\dimen0<\pagegoal
     \dimen0=\ht0\advance\dimen0 by\dp0\advance\dimen0 by
     3\normalbaselineskip
     \advance\dimen0 by\pagetotal
     \ifdim\dimen0>\pagegoal\eject\fi\fi
     \noindent
     \if!#1!\ignorespaces#2
     \else\setbox0=\hbox{\ignorespaces#1\unskip\ }\hangindent=\wd0
     \hangafter=1\box0\ignorespaces#2\fi
     \vskip18pt plus4pt minus4pt\egroup
     \nobreak
     \parindent=0pt
     \everypar={\global\parindent=\oldparindent
     \global\let\lasttitle=N\global\everypar={}}%
     \global\let\lasttitle=A%
     \ignorespaces}
 \def\titleb#1#2{\if N\lasttitle\else\vskip-28pt
     \fi
     \vskip18pt plus 4pt minus4pt
     \bgroup
\textfont0=\tenbf \scriptfont0=\sevenbf \scriptscriptfont0=\fivebf
\textfont1=\tams \scriptfont1=\tamss \scriptscriptfont1=\tbmss
     \lineskip=0pt
     \pretolerance=10000
     \noindent
     \bf
     \rightskip 0pt plus 6em
     \setbox0=\vbox{\vskip23pt\def\fonote##1{}%
     \noindent
     \if!#1!\ignorespaces#2
     \else\setbox0=\hbox{\ignorespaces#1\unskip\enspace}\hangindent=\wd0
     \hangafter=1\box0\ignorespaces#2\fi
     \vskip10pt}%
     \dimen0=\pagetotal\advance\dimen0 by-\pageshrink
     \ifdim\dimen0<\pagegoal
     \dimen0=\ht0\advance\dimen0 by\dp0\advance\dimen0 by
     3\normalbaselineskip
     \advance\dimen0 by\pagetotal
     \ifdim\dimen0>\pagegoal\eject\fi\fi
     \noindent
     \if!#1!\ignorespaces#2
     \else\setbox0=\hbox{\ignorespaces#1\unskip\enspace}\hangindent=\wd0
     \hangafter=1\box0\ignorespaces#2\fi
     \vskip8pt plus4pt minus4pt\egroup
     \nobreak
     \parindent=0pt
     \everypar={\global\parindent=\oldparindent
     \global\let\lasttitle=N\global\everypar={}}%
     \global\let\lasttitle=B%
     \ignorespaces}
 \def\titlec#1#2{\if N\lasttitle\else\vskip-23pt
     \fi
     \vskip18pt plus 4pt minus4pt
     \bgroup
\textfont0=\tenbfne \scriptfont0=\sevenbf \scriptscriptfont0=\fivebf
\textfont1=\tams \scriptfont1=\tamss \scriptscriptfont1=\tbmss
     \tenbfne
     \lineskip=0pt
     \pretolerance=10000
     \noindent
     \rightskip 0pt plus 6em
     \setbox0=\vbox{\vskip23pt\def\fonote##1{}%
     \noindent
     \if!#1!\ignorespaces#2
     \else\setbox0=\hbox{\ignorespaces#1\unskip\enspace}\hangindent=\wd0
     \hangafter=1\box0\ignorespaces#2\fi
     \vskip6pt}%
     \dimen0=\pagetotal\advance\dimen0 by-\pageshrink
     \ifdim\dimen0<\pagegoal
     \dimen0=\ht0\advance\dimen0 by\dp0\advance\dimen0 by
     2\normalbaselineskip
     \advance\dimen0 by\pagetotal
     \ifdim\dimen0>\pagegoal\eject\fi\fi
     \noindent
     \if!#1!\ignorespaces#2
     \else\setbox0=\hbox{\ignorespaces#1\unskip\enspace}\hangindent=\wd0
     \hangafter=1\box0\ignorespaces#2\fi
     \vskip6pt plus4pt minus4pt\egroup
     \nobreak
     \parindent=0pt
     \everypar={\global\parindent=\oldparindent
     \global\let\lasttitle=N\global\everypar={}}%
     \global\let\lasttitle=C%
     \ignorespaces}
 \def\titled#1{\if N\lasttitle\else\vskip-\baselineskip
     \fi
     \vskip12pt plus 4pt minus 4pt
     \bgroup
\textfont1=\tams \scriptfont1=\tamss \scriptscriptfont1=\tbmss
     \bf
     \noindent
     \ignorespaces#1\ \ignorespaces\egroup
     \ignorespaces}
\let\ts=\thinspace
\def\footnoterule{\kern-3pt\hrule width 1.66666 cm\kern2.6pt}
\newcount\footcount \footcount=0
\def\advftncnt{\advance\footcount by1\global\footcount=\footcount}
\def\fonote#1{\advftncnt$^{\the\footcount}$\begingroup\petit
\parfillskip=0pt plus 1fil
\def\textindent##1{\hangindent0.5\oldparindent\noindent\hbox
to0.5\oldparindent{##1\hss}\ignorespaces}%
\vfootnote{$^{\the\footcount}$}{#1\vskip-9.69pt}\endgroup}
\def\item#1{\par\noindent
\hangindent6.5 mm\hangafter=0
\llap{#1\enspace}\ignorespaces}

\def\titleao#1{\vfill\supereject
\ifodd\pageno\else\null\vfill\supereject\fi
\let\header=N
     \bgroup
\textfont0=\tafontt \scriptfont0=\tafonts \scriptscriptfont0=\tafontss
\textfont1=\tamt \scriptfont1=\tams \scriptscriptfont1=\tamss
\textfont2=\tast \scriptfont2=\tass \scriptscriptfont2=\tasss
\textfont3=\tasyt \scriptfont3=\tasys \scriptscriptfont3=\tenex
     \baselineskip=18pt
     \lineskip=0pt
     \pretolerance=10000
     \tafontt
     \centerpar{#1}%
     \vskip75pt\egroup
     \nobreak
     \parindent=0pt
     \everypar={\global\parindent=\oldparindent
     \global\let\lasttitle=N\global\everypar={}}%
     \global\let\lasttitle=A%
     \ignorespaces}






\def\leaderfill{\kern0.5em\leaders\hbox to 0.5em{\hss.\hss}\hfill\kern
0.5em}
\newdimen\chapindent
\newdimen\sectindent
\newdimen\subsecindent
\newdimen\thousand
\setbox0=\hbox{\bf 10. }\chapindent=\wd0
\setbox0=\hbox{10.10 }\sectindent=\wd0
\setbox0=\hbox{10.10.1 }\subsecindent=\wd0
\setbox0=\hbox{\enspace 100}\thousand=\wd0
\def\contpart#1#2{\medskip\noindent
\vbox{\kern10pt\leftline{\textfont1=\tams
\scriptfont1=\tamss\scriptscriptfont1=\tbmss\bf
\advance\chapindent by\sectindent
\hbox to\chapindent{\ignorespaces#1\hss}\ignorespaces#2}\kern8pt}%
\let\lasttitle=Y\par}
\def\contcontribution#1#2{\if N\lasttitle\bigskip\fi
\let\lasttitle=N\line{{\textfont1=\tams
\scriptfont1=\tamss\scriptscriptfont1=\tbmss\bf#1}%
\if!#2!\hfill\else\leaderfill\hbox to\thousand{\hss#2}\fi}\par}
\def\conttitlea#1#2#3{\line{\hbox to
\chapindent{\strut\bf#1\hss}{\textfont1=\tams
\scriptfont1=\tamss\scriptscriptfont1=\tbmss\bf#2}%
\if!#3!\hfill\else\leaderfill\hbox to\thousand{\hss#3}\fi}\par}
\def\conttitleb#1#2#3{\line{\kern\chapindent\hbox
to\sectindent{\strut#1\hss}{#2}%
\if!#3!\hfill\else\leaderfill\hbox to\thousand{\hss#3}\fi}\par}
\def\conttitlec#1#2#3{\line{\kern\chapindent\kern\sectindent
\hbox to\subsecindent{\strut#1\hss}{#2}%
\if!#3!\hfill\else\leaderfill\hbox to\thousand{\hss#3}\fi}\par}
\long\def\lemma#1#2{\removelastskip\vskip\baselineskip\noindent{\tenbfne
Lemma\if!#1!\else\ #1\fi\ \ }{\it\ignorespaces#2}\vskip\baselineskip}
\long\def\proposition#1#2{\removelastskip\vskip\baselineskip\noindent{\tenbfne
Proposition\if!#1!\else\ #1\fi\ \ }{\it\ignorespaces#2}\vskip\baselineskip}
\long\def\theorem#1#2{\removelastskip\vskip\baselineskip\noindent{\tenbfne
Theorem\if!#1!\else\ #1\fi\ \ }{\it\ignorespaces#2}\vskip\baselineskip}
\long\def\corollary#1#2{\removelastskip\vskip\baselineskip\noindent{\tenbfne
Corollary\if!#1!\else\ #1\fi\ \ }{\it\ignorespaces#2}\vskip\baselineskip}
\long\def\example#1#2{\removelastskip\vskip\baselineskip\noindent{\tenbfne
Example\if!#1!\else\ #1\fi\ \ }\ignorespaces#2\vskip\baselineskip}
\long\def\exercise#1#2{\removelastskip\vskip\baselineskip\noindent{\tenbfne
Exercise\if!#1!\else\ #1\fi\ \ }\ignorespaces#2\vskip\baselineskip}
\long\def\problem#1#2{\removelastskip\vskip\baselineskip\noindent{\tenbfne
Problem\if!#1!\else\ #1\fi\ \ }\ignorespaces#2\vskip\baselineskip}
\long\def\solution#1#2{\removelastskip\vskip\baselineskip\noindent{\tenbfne
Solution\if!#1!\else\ #1\fi\ \ }\ignorespaces#2\vskip\baselineskip}


\long\def\definition#1#2{\removelastskip\vskip\baselineskip\noindent{\tenbfne
Definition\if!#1!\else\
#1\fi\ \ }\ignorespaces#2\vskip\baselineskip}
\def\frame#1{\bigskip\vbox{\hrule\hbox{\vrule\kern5pt
\vbox{\kern5pt\advance\hsize by-10.8pt
\centerline{\vbox{#1}}\kern5pt}\kern5pt\vrule}\hrule}\bigskip}
\def\frameddisplay#1#2{$$\vcenter{\hrule\hbox{\vrule\kern5pt
\vbox{\kern5pt\hbox{$\displaystyle#1$}%
\kern5pt}\kern5pt\vrule}\hrule}\eqno#2$$}
\def\typeset{\petit\noindent This book was processed by the author using
the \TeX\ macro package from Springer-Verlag.\par}
\outer\def\byebye{\bigskip\bigskip\typeset
\footcount=1\ifx\speciali\undefined\else
\loop\smallskip\noindent special character No\number\footcount:
\csname special\romannumeral\footcount\endcsname
\advance\footcount by 1\global\footcount=\footcount
\ifnum\footcount<11\repeat\fi
\gdef\leftheadline{\hbox to2.08333 cc{\folio\hfil}\ignorespaces
\the\AUTHOR\unskip: \the\HEAD\hfill}\vfill\supereject\end}